\documentclass[conference]{IEEEtran}
\usepackage{algorithm}
\usepackage[noend]{algpseudocode}

\usepackage{amsmath}
\usepackage{mathtools}
\usepackage{amsmath,amsfonts,amssymb}
\usepackage{booktabs}
\usepackage{mathrsfs}
\usepackage{makecell}
\usepackage{multirow}
\usepackage{xcolor}
\usepackage[noadjust]{cite}

\usepackage[switch,columnwise]{lineno}
\usepackage{amssymb}
\usepackage{pifont}
\usepackage[hang]{footmisc}

\begin{document}
\bstctlcite{IEEEexample:BSTcontrol}
\title{Closed-Loop Neural Interfaces \\with Embedded Machine Learning}


\vspace{-8mm}
\author{
Bingzhao~Zhu$^{1,*}$, Uisub~Shin$^{1,*}$, Mahsa Shoaran$^2$ 
\\$^1$School of Electrical and Computer Engineering, Cornell University, Ithaca, NY, USA\\
$^2$Institute of Electrical Engineering \& Center for Neuroprosthetics, EPFL, 1202 Geneva, Switzerland\\
Email: bz323@cornell.edu; us52@cornell.edu; mahsa.shoaran@epfl.ch 
\\$^*$These authors contributed equally to this work.
\vspace{-8mm}
}


\maketitle
\setlength{\skip\footins}{2pt}
\renewcommand\footnotemargin{0.00001pt}

\begin{abstract}
Neural interfaces capable of multi-site electrical recording, on-site signal classification, and closed-loop therapy are critical for the diagnosis and treatment of neurological disorders. However, deploying machine learning algorithms on low-power neural devices is challenging, given the tight constraints on computational and memory resources for such devices. In this paper, we review the recent developments in embedding machine learning in neural interfaces, with a focus on design trade-offs and hardware efficiency. We also present our optimized tree-based model for low-power and memory-efficient classification of neural signal in brain implants. Using energy-aware learning and model compression, we show that the proposed oblique trees can outperform conventional machine learning models in applications such as seizure or tremor detection and motor decoding. 
\end{abstract}


%
\IEEEpeerreviewmaketitle

 \vspace{-4mm}
\section{Introduction}  
\begin{figure}[t]
  \centering
  \includegraphics[width=0.8\columnwidth]{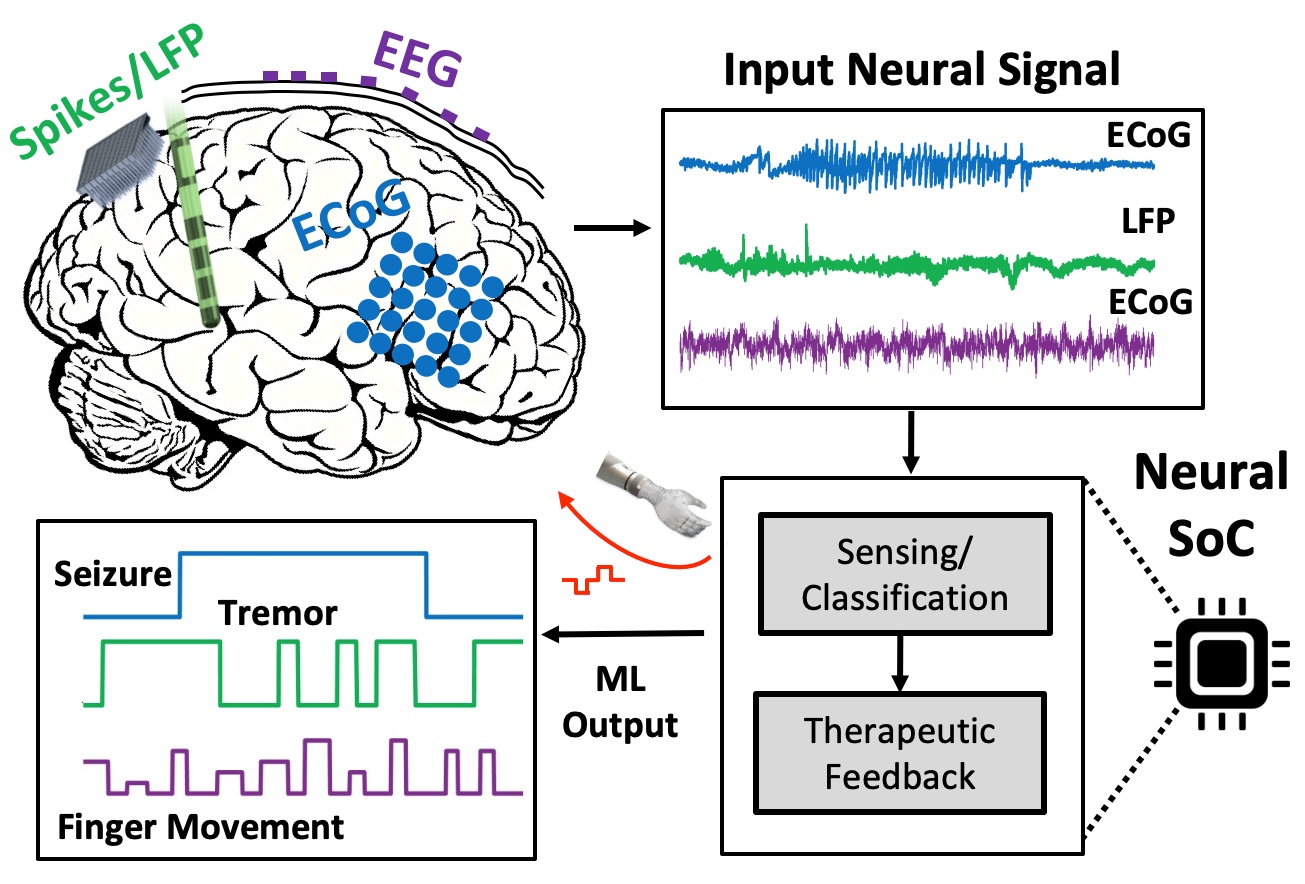}\vspace{-3mm}
  \caption{Block diagram of a closed-loop neural interface; Neural signals are recorded with invasive  or noninvasive electrodes, and an embedded classifier  detects disease symptoms or decodes a movement. Closed-loop feedback is enabled to suppress an abnormal activity or control a prosthetic device.}\vspace{-5mm}
  \label{block}
\end{figure}

%
%
%
%
\IEEEPARstart{C}\noindent losed-loop neural interfaces enhance the therapeutic efficacy while alleviating side effects and improving the battery life in their open-loop counterparts. For example, adaptive stimulation in response to abnormal brain activity has shown promise in treating epileptic seizures \cite{morrell2011responsive} and motor symptoms of Parkinson’s disease \cite{little2013adaptive}. Recently, the application of machine learning (ML)  in closed-loop neural interfaces and its ASIC implementation has gained interest among researchers. Real-time neural processing can be enabled through on-chip feature extraction and classification, followed by a closed-loop feedback (e.g., neurostimulation) to suppress a symptom, provide a sensory feedback, or control a prosthetic device in a brain-machine interface (BMI), as illustrated in Fig. \ref{block}. The ASIC realization of ML is particularly favorable in such implants, enabling real-time near-sensor processing, triggering a therapeutic feedback, lowering the data transmission rate, and alleviating security and privacy concerns. 

Despite the promise and benefits of machine learning for closed-loop neural interfacing, stringent energy and area constraints on multi-channel neural implants pose significant challenges for the ASIC implementation of ML models. Therefore, it is critical to develop  hardware-efficient ML solutions to overcome such limitations. This paper reviews the state-of-the-art neural interfaces with embedded classification 
and describes several techniques for energy, area  and memory efficiency, including single-path inference, cost-aware learning, and model compression. A new class of oblique tree-based models suited for hardware-efficient realization is proposed and verified on three neural signal classification tasks. 

\section{Neural Interfaces with Embedded ML}
Various machine learning algorithms and hardware architectures have been reported for  neurological disease detection \cite{lee2013low,altaf20131,shoaran2018energy,shoaran2016hardware,o202026,chen2014fully,yao2020improved,zhu2019migraine, fang2019development, aslam2020a10, chang2019ultra} and brain-machine interfacing \cite{chen2015128, yang2017hardware, do2018area} using either invasive or noninvasive electrodes, as summarized below. 

\begin{figure*}[t]
  \centering
  \includegraphics[width=2\columnwidth]{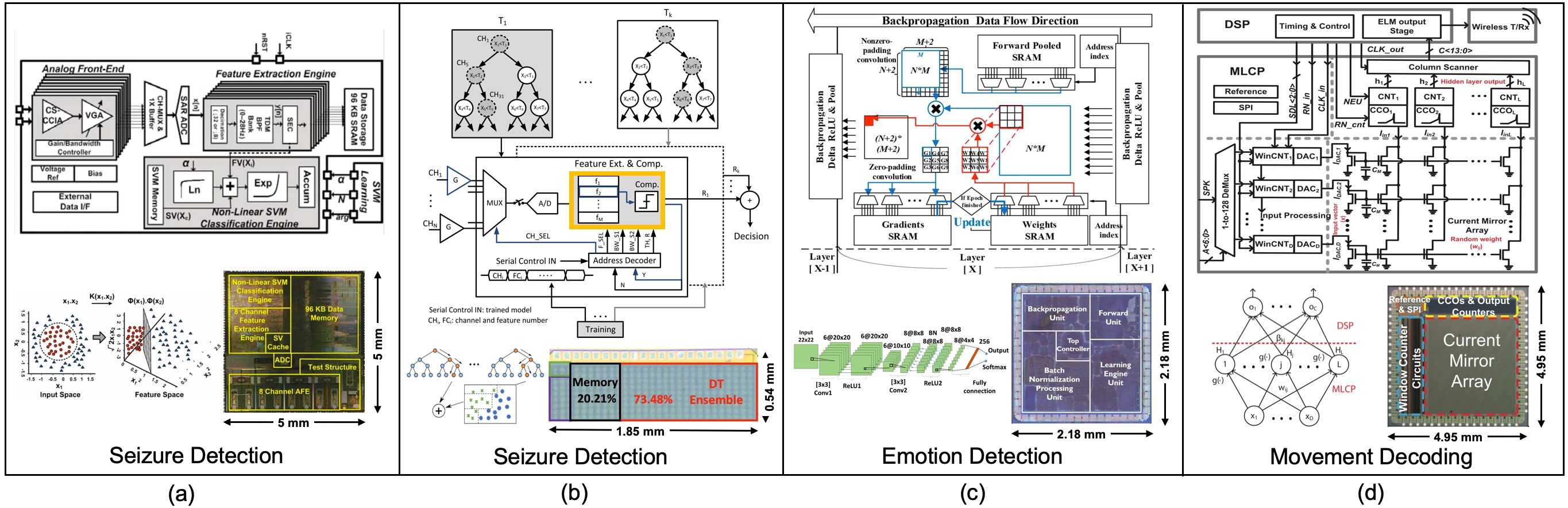}
  \vspace{-3mm}
  \caption{Hardware block diagram and chip micrograph of learning models for different applications: (a) non-linear SVM for seizure detection \cite{altaf20131}, (b) gradient-boosted trees for seizure detection \cite{shoaran2018energy}, (c) CNN for emotion detection \cite{fang2019development}, and (d) ELM for motor intention decoding \cite{chen2015128}.}
  \label{fig2}
\end{figure*}
\vspace{-4mm}
\begin{table*}[t]\vspace{-3mm}
\centering
\caption{Comparison of Machine Learning SoCs}
\vspace{-3mm}
\scalebox{0.75}{
\begin{tabular}{l|c|c|c|c|c|c|c|c|c}
\hline \hline
Parameter & JETCAS'18 \cite{shoaran2018energy} & ISSCC'13 \cite{altaf20131} & JSSC'13 \cite{chen2014fully} & ISSCC'20 \cite{o202026} & JETCAS'19 \cite{fang2019development} & TCAS-I'19 \cite{chang2019ultra} & TBioCAS'16 \cite{chen2015128} & TBioCAS'17 \cite{yang2017hardware} & TVLSI'19 \cite{do2018area}\\
\hline \hline
Process & 65 nm & 180 nm & 180 nm & 65 nm & 28 nm & 180 nm & 350 nm & 130 nm$^\dag$$^\dag$ & 65 nm\\
\hline
Classifier & XGB DT & Non-Lin SVM & LLS & AdaBoost DT & CNN & NN-based DT & ELM & DT & K-means\\
\hline
Application & Seizure Det. & Seizure Det. & Seizure Det. & Seizure Det. & Emotion Det. & Sleep Staging & Motor Decoding & Spike Sorting & Spike Sorting\\
\hline
Signal Modality & iEEG & EEG & iEEG & iEEG & EEG & EEG, EMG & Monkey LFP & Rat LFP & Synthetic Spikes\\
\hline
No. of Channels & 32 & 8 & 8 & 8 & 6 & 2 & 128 & 32 & 128\\
\hline
Energy Eff. (or Power) & 41.2 nJ/class. & 1.22 $\mu$J/class.$^\ast$$^\ast$ & 77.9 $\mu$J/class. & 36 nJ/class. & 76.61 mW & 0.149 mJ/epoch & 16.2 nJ/class. & 24 $\mu$W & 22.4 $\mu$W\\
\hline
Memory & 1 kB & N.A. & N.A. & N.A. & 11.8 kB & 6.4 kB & N.A. & 0.56 kB & 4.88 kB\\
\hline
Area$^\ast$ & 0.8 mm$^2$ & 5.63 mm$^2$ & 4.85 mm$^2$ & 0.71 mm$^2$ & 3.47 mm$^2$ & 8.77 mm$^2$ & 17.4 mm$^2$ & 0.73 mm$^2$ & 0.41 mm$^2$\\
\hline
Sensitivity & 83.7\% & 95.1\% & 92\% & 96.7\% & 83.4\%$^\ddag$ & 81\%$^\ddag$ & 99.3\%$^\ddag$ & $\sim$77\%$^\ddag$ & 72/86\%$^\ddag$$^\S$\\
\hline
Specificity & 88.1\% & 0.27 FA/h$^\dag$ & N.A. & 0.8 FA/h$^\dag$ & N.A. & N.A. & N.A. & N.A. & N.A.\\
\hline
Latency & 1.79 s & 2 s & 0.8 s & N.A. & 0.45 s & N.A. & N.A. & N.A. & N.A.\\
\hline \hline
\multicolumn{5}{l}{$\ast$ Estimated area of feature extractor and classifier from chip micrographs (excluding pads) } &
\multicolumn{5}{l}{$\ast\ast$ Estimated from power breakdown} \\
\multicolumn{5}{l}{$\dag$ Number of false alarms per hour} &
\multicolumn{5}{l}{$\dag\dag$ Post-synthesis results} \\
\multicolumn{5}{l}{$\ddag$ Accuracy metric} &
\multicolumn{5}{l}{$\S$ For unsupervised and semisupervised modes, respectively}\vspace{-6mm}
\end{tabular} 
\label{comparison} }
\end{table*}

\vspace{2mm}
\subsection{Symptom Detection}  
Accurate detection of symptoms in neurological disorders is the first step toward an effective closed-loop stimulation therapy. A typical example is epileptic seizure detection, where a supervised machine learning algorithm could be used to detect \textit{seizure} events from electrophysiological recordings. The majority of seizure detection SoCs in literature have adopted support vector machine (SVM) classifiers \cite{lee2013low, altaf20131}, Fig.~\ref{fig2}(a). SVMs typically require a large number of multiply-and-accumulate (MAC) and non-linear operations, while their computational and memory resources linearly scale with the number of channels and input features. 

Recently, ensembles of decision trees (DTs) such as gradient boosting trees and random forests  have emerged as an accurate yet hardware-friendly  solution for resource-constrained platforms. With simple comparisons applied to input features, tree-based models enable low-complexity hardware architectures. 
In \cite{shoaran2018energy}, an ensemble of eight gradient-boosted DTs achieved an energy efficiency of 41.2nJ/class for 32-channel intracranial EEG (iEEG)-based seizure detection, Fig.~\ref{fig2}(b).
A sequential feature extraction approach enabled the use of a single  feature extraction engine  per tree, thus significantly reducing the hardware cost for multi-channel implants. 
Another tree-based on-chip classifier was recently reported  for epileptic seizure detection~\cite{o202026}, where 1024  decision stumps (i.e., trees of depth one) were aggregated using  AdaBoost technique. 
With bit-serial operation and on-chip weight regeneration, the 8-channel SoC reported an energy efficiency of 36nJ/class.
Moreover, DT ensembles have shown superior performance in other neural tasks such as Parkinsonian tremor detection using local field potentials (LFP) \cite{yao2020improved, yao2018resting} and migraine state classification from somatosensory evoked potentials (SSEP)~\cite{zhu2019migraine}. 
ML models have also been explored for neural signal classification in applications such as emotion detection \cite{fang2019development, aslam2020a10}, sleep stage classification \cite{chang2019ultra}, and for predicting memory dysfunction~\cite{ezzyat2018closed} and mental fatigue~\cite{yao2020mental} (to potentially trigger a neurostimulation therapy). 
In \cite{fang2019development}, a convolutional neural network (CNN) SoC with online training capability was implemented for emotion recognition, Fig.~\ref{fig2}(c). To reduce the memory and area utilized by batch processing,  training and acceleration were executed in four phases through re-using minibatch data and hardware, at the cost of increased training time. Combined with an external feature extraction processor, the CNN classifier achieved an accuracy of 83.36\%  in a binary emotion detection task. A 4-layer  neural network classifier was recently reported for  emotion detection in autistic children~\cite{aslam2020a10}. This 2-channel EEG processor achieved a classification accuracy of 85.2\% while consuming 10.1$\mu$J/class. 
\vspace{-2mm}
\subsection{Brain-Machine Interfaces}\vspace{-1mm}
BMIs provide a communication channel between the human brain and external environment  for paralyzed patients. Similar to implants for disease detection, BMIs also operate on a resource-constrained platform, making it crucial to design hardware-friendly  movement decoders for fully implantable BMIs. A variety of signal modalities  such as EEG, ECoG, spikes and LFP can be used as input to a BMI, providing various degrees of motor control and invasiveness.
An intracortical decoder based on extreme learning machine (ELM) was reported in \cite{chen2015128}, exploiting the CMOS device mismatch for random weight generation and sub-threshold current-mode operation, as shown in Fig.~\ref{fig2}(d). 
The ELM processor combined with a DSP (performing additional MAC operations and spike sorting) consumed 16.2nJ/class.
A scalable DT-based spike sorting processor for high-channel-count BMIs was reported in \cite{yang2017hardware}, 
while \cite{do2018area} presents an area-efficient spike sorting processor with online K-means clustering. 
The hardware efficiency of the processor was improved using  multiplier-less spike detection and feature extraction, time-multiplexed registers, and low-voltage SRAM. 

The hardware cost and performance of state-of-the-art classifiers in neural interface SoCs are summarized in Table~\ref{comparison}. 
As shown in this table, DTs provide an attractive solution for low-power and area-limited applications. 
It should be noted that when comparing different ML SoCs, various factors such as classification task, channel count, feature type,  and signal modality need to be taken into account. A comparison of different  classifiers for  iEEG and EEG-based seizure detection can be found in \cite{shoaran2018energy}, \cite{LR}, while the hardware complexity of DTs and deep neural networks is compared in~\cite{taghavi2019hardware}.  
\vspace{-1mm}
\section{Hardware-Algorithm Optimized DTs} \vspace{-1mm}
Compared to the conventional approach of transmitting raw neural data for off-the-body  processing, neural implants with embedded ML avoid the use of power-demanding transmitters. Yet, efficient implementation of ML  is pivotal to minimize heat dissipation and battery usage. Moreover, small area of the implant and large number of channels require the use of minimal silicon and memory resources. On-chip classifiers enable a fast inference and  provide a timely feedback to the patients, leading to a rapid activation of therapeutic stimulation or prosthetic control. 
Overall, embedded ML models are required to consume low power and small area, while  providing a low detection latency and high classification accuracy. 
\begin{figure}[t]
  \centering \vspace{-4mm}
  \includegraphics[width=1\columnwidth]{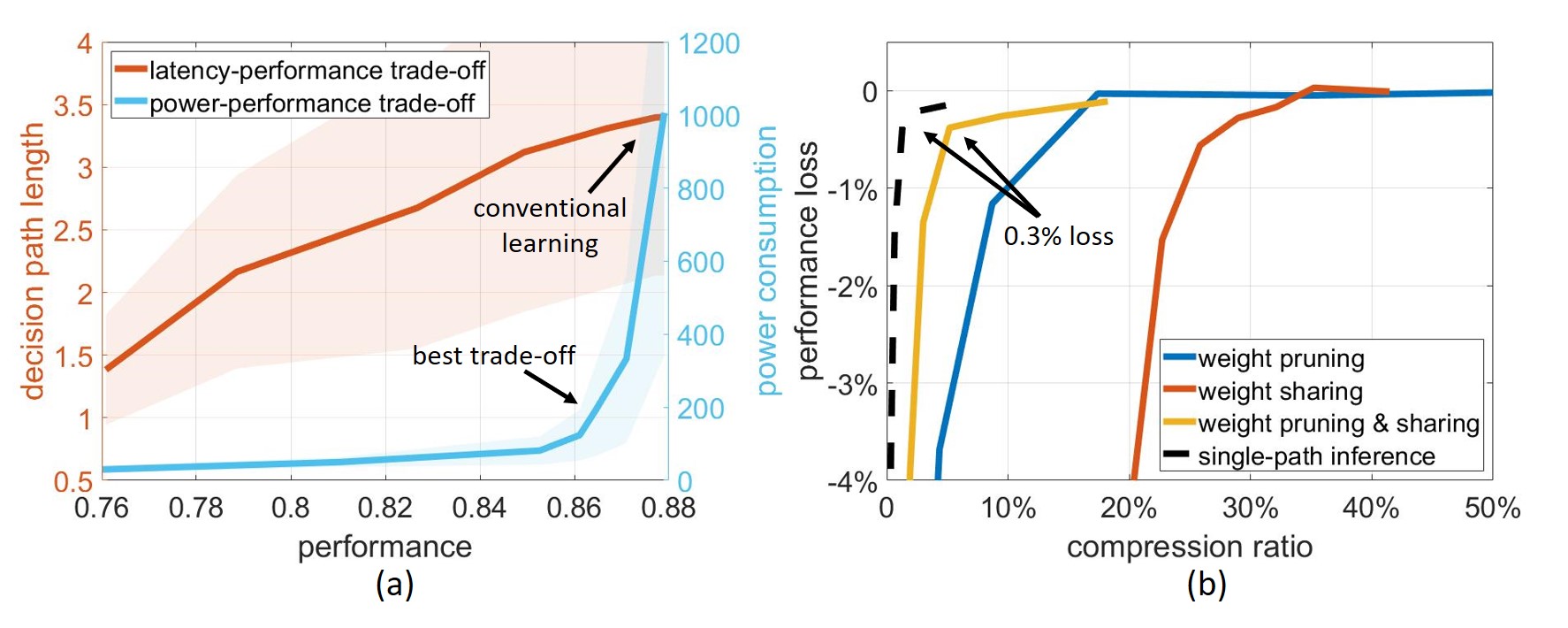}
  \vspace{-8mm}
  \caption{(a) Trade-off between classification performance and latency/power for seizure detection. The inference power  is normalized to the power of most efficient feature (line-length). Shading areas indicate standard deviation across patients; (b) Model compression with weight pruning and sharing. 
  In the single-path scheme, only the parameters used during inference are included. 
  This experiment was conducted for an oblique tree on the MNIST dataset.}
  \label{fig3n4}
  \vspace{-5mm}
\end{figure}

Among the ML algorithms widely used in embedded neural SoCs, tree-based models are compatible with a single-path inference scheme \cite{shoaran2018energy, zhu2020resot} where a single root-to-leaf path is visited to make predictions (i.e., a small portion of the model). 
This lightweight inference  is a significant advantage, considering the large number of channels in modern neural interface platforms. 
DTs can be further optimized for power- and memory-efficient implementation through a wise co-design of algorithm and hardware, as described below. 
\vspace{-1mm}
\subsection{Cost-Aware Learning}\vspace{-1mm}
During inference, the major hardware cost (e.g., power, area or latency) of tree-based models is associated with feature extraction \cite{shoaran2018energy}. As the inference time increases proportional to the length of a decision path, sequential processing may raise a concern on detection latency.
In \cite{shoaran2018energy}, we used an asynchronous approach  to reduce the latency due to  single-path sequential processing. A cost-aware learning scheme may reduce the power and/or latency by incorporating these cost factors into  objective function as a regularization term \cite{zhu2019cost, zhu2019hardware}: $\min  \sum_{i} L\left(y_{i}, f\left(\boldsymbol{x}_{i}\right)\right)+\lambda \Psi\left(f, \boldsymbol{x}_{\boldsymbol{i}}\right)$.
Here, we seek to learn a decision function $f(x)$ that minimizes the loss $L\left(y_{i}, f\left(\boldsymbol{x}_{i}\right)\right)$ in conjunction with the computational cost $\Psi\left(f, \boldsymbol{x}_{\boldsymbol{i}}\right)$. As an example, we study the impact of optimizing the model for power and latency. The power consumption for extracting  various features was estimated for a standard digital implementation in 65nm CMOS process \cite{zhu2020resot}. Taking seizure detection task as an example, the line-length feature ($\frac{1}{d} \sum_{d}|x[n]-x[n-1]|$, $d=$ window size) consumes a negligible power, whereas band power features are more power demanding due to the FIR filtering stage. For latency estimation, we used the length of the decision path, which is upper bounded by the maximum depth of the tree.

\begin{figure}[t]
  \centering
  \vspace{-4mm}
  \includegraphics[width=0.8\columnwidth]{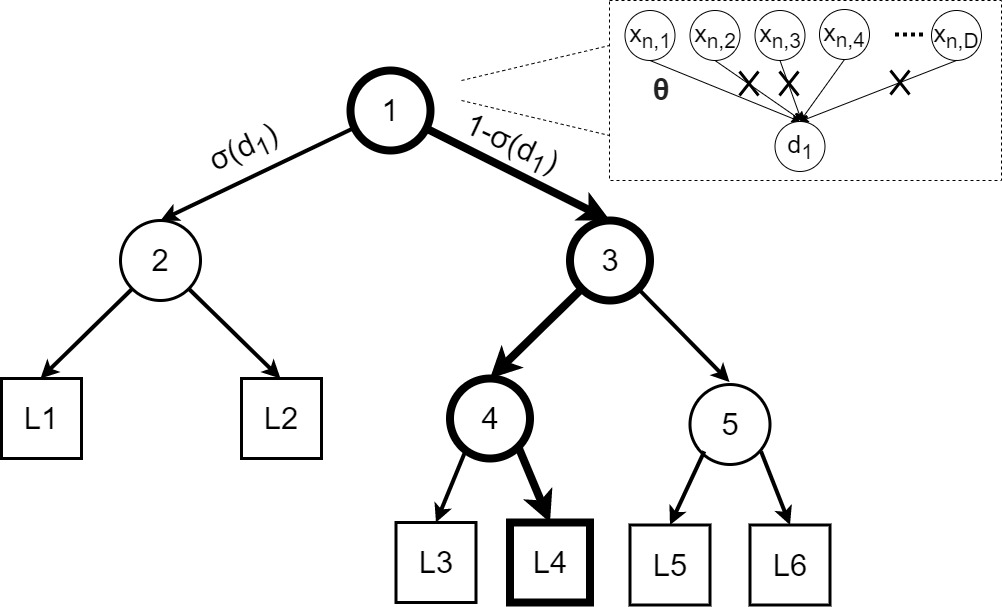}
  \vspace{-2mm}
  \caption{
  Proposed oblique tree with probabilistic routing. In the inference phase, the test samples follow the most probable path. Inside internal nodes, the decision functions are represented by a two-layer neural network \cite{zhu2020resot}.}\vspace{-5mm}
  \label{neurtree}
\end{figure}

Figure \ref{fig3n4}(a) shows the trade-off between classification performance (F1 score) and cost metrics (latency, power). An ensemble of axis-aligned trees was used to detect epileptic seizures  in 10 patients \cite{zhu2020resot}. The number of trees and maximum depth were optimized using 5-fold cross-validation on each subject. As shown in Fig. \ref{fig3n4}(a), the classification performance degrades by shortening the decision path, thus requiring an asynchronous learning scheme  to reduce the latency \cite{shoaran2018energy}. On the other hand, a power-efficient performance region is observed in Fig. \ref{fig3n4}(a), where we can drastically reduce the power consumption with only a marginal performance loss.
\vspace{-1mm}
\subsection{Model Compression}\vspace{-1mm}
Compression techniques such as fixed-point quantization, weight pruning and sharing have been widely used to reduce the model size in DNNs \cite{han2015deep,lin2016fixed}. Similar techniques can be used to develop hardware-friendly DTs that are  more efficiently deployable on ASIC.
Importantly, conventional tree ensembles may suffer from a large model size due to the large number of trees required in non-trivial  classification tasks \cite{zhu2019migraine,  o202026, yao2019enhanced}. To partially alleviate this issue, we quantized the tree parameters to reduce  model size and allow fixed-point arithmetic \cite{zhu2019hardware, zhu2020resot}. 
In a boosting framework, threshold values and leaf weights were quantized with 10 and 3 bits for  seizure detection, reducing the model size by 2.4$\times$ compared to a gradient boosting ensemble with floating point weights.

\begin{figure}[t]
  \centering
  \includegraphics[width=1\columnwidth]{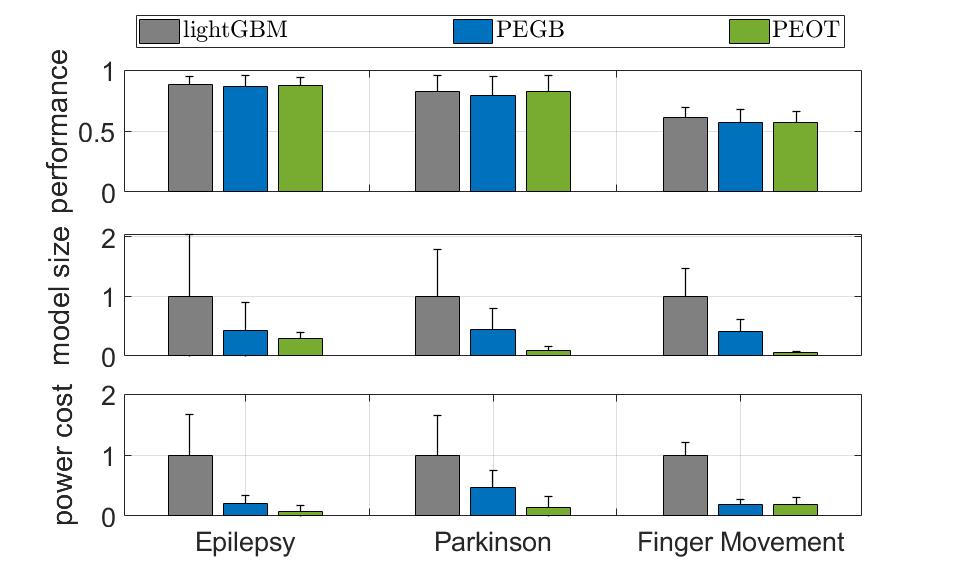}
  \vspace{-3mm}
  \caption{Comparison of tree-based models on three neural tasks. Ensemble of gradient boosted trees (lightGBM \cite{ke2017lightgbm}), power-efficient gradient boosted trees with fixed-point quantization (PEGB \cite{zhu2019cost}), and power-efficient oblique trees with weight pruning and sharing (PEOT \cite{zhu2020resot}) were compared. Average model sizes and power costs are normalized to the performance of lightGBM. 
  Error bars represent the standard deviation among subjects.
  }\vspace{-4mm}
  \label{fig5}
\end{figure}

Unlike axis-aligned decision trees, oblique  trees involve multiple features in their internal nodes and can generate accurate predictions with a single tree (Fig.~\ref{neurtree}) \cite{zhu2020resot}. Interestingly, within a probabilistic routing scheme, oblique trees can be trained using gradient-based optimization, similar to a neural network. 
Here, the decision functions in the internal nodes can be represented by a two-layer neural network, for which weight pruning and sharing techniques can be used to create sparse connections.
Figure \ref{fig3n4}(b) illustrates the trade-off between classification performance and model size for an oblique tree (OT) with a maximum depth of 4, trained  on the MNIST dataset. With weight pruning and sharing, the model size was compressed by 20$\times$ with only a marginal performance loss (0.3\%). Moreover, with single-path inference only a small portion (26.3\%) of the parameters were used, improving the  hardware efficiency during inference. Combining  compression with cost-aware learning, we built power-efficient oblique trees (PEOT) and benchmarked them against conventional (lightGBM \cite{ke2017lightgbm, shoaran2018energy}) and power-efficient gradient boosted trees (PEGB \cite{zhu2019cost}). Testing on three different neural tasks including seizure detection (iEEG, 10 patients),  tremor detection (LFP, 12 patients), and finger movement classification (ECoG, 9 subjects),  PEOT reduced the model sizes by 10.5$\times$ and the power cost by 8.8$\times$, Fig.~\ref{fig5}. The PEOT  model also achieved average reduction factors of 4.4$\times$ in model size and 2.3$\times$ in power cost compared to PEGB.

\vspace{-1mm}
\section{Conclusion}\vspace{-1mm}
We reviewed a recent trend in the development of closed-loop neural interfaces that embed ML on chip. Algorithm and hardware approaches for ML SoCs  in various neural applications were discussed. 
We proposed a power-efficient oblique tree model which integrates cost-aware learning, weight pruning and sharing. Testing on three neural classification tasks, the proposed  model improved the energy and memory efficiency while maintaining the classification performance. 


\ifCLASSOPTIONcaptionsoff
  \newpage
\fi



%

\vspace{-1mm}
\bibliographystyle{IEEEtran}
\vspace{-1mm}
\bibliography{cit}

\end{document}